\documentclass[12pt,preprint]{emulateapj}

\shorttitle{AGN/SF Diagnostics at $z>1$}
\shortauthors{Trump et al.}

\begin{document}

\title{Testing Diagnostics of Nuclear Activity and Star Formation in
  Galaxies at $z>1$\altaffilmark{*}}

\author{
  Jonathan R. Trump,\altaffilmark{1}
  Nicholas P. Konidaris,\altaffilmark{2}
  Guillermo Barro,\altaffilmark{1}
  David C. Koo,\altaffilmark{1}
  Dale D. Kocevski,\altaffilmark{3}
  St\'{e}phanie Juneau,\altaffilmark{5}
  Benjamin J. Weiner,\altaffilmark{4}
  S. M. Faber,\altaffilmark{1}
  Ian S. McLean,\altaffilmark{6}
  Renbin Yan,\altaffilmark{3}
  Pablo G. P\'{e}rez-Gonz\'{a}lez,\altaffilmark{7}
  and Victor Villar\altaffilmark{7}
}

\altaffiltext{*}{
  Based on observations with the NASA/ESA \emph{Hubble Space
  Telescope}, obtained at the Space Telescope Science Institute, which
  is operated by AURA Inc, under NASA contract NAS 5-26555.  Also
  based on data obtained at the W.~M.~Keck Observatory, made possible
  by the generous financial support of the W.~M.~Keck Foundation and
  operated as a scientific partnership among Caltech, the University
  of California, and NASA.
\label{candels}}

\altaffiltext{1}{
  University of California Observatories/Lick Observatory and
  Department of Astronomy and Astrophysics, University of California,
  Santa Cruz, CA 95064 USA
\label{UCO/Lick}}

\altaffiltext{2}{
  California Institute of Technology, MC 105-24, 1200 East California
  Boulevard, Pasadena, CA 91125 USA
\label{Caltech}}

\altaffiltext{3}{
  Department of Physics and Astronomy, University of Kentucky,
  Lexington, KY 40506
\label{Kentucky}}

\altaffiltext{4}{
  Steward Observatory, University of Arizona, 933 North Cherry Avenue,
  Tucson, AZ 85721 USA
\label{Arizona}}

\altaffiltext{5}{
  Irfu/Service d'Astrophysique, CEA-Saclay, Orme des Merisiers, 91191
  Gif-sur-Yvette Cedex, France
\label{Saclay}}

\altaffiltext{6}{
  Department of Physics and Astronomy, UCLA, Los Angeles, CA 90095, USA
\label{UCLA}}

\altaffiltext{7}{
  Departamento de Astrof\'{i}sica, Facultad de CC. F\'{i}sicas,
  Universidad Complutense de Madrid, E-28040 Madrid, Spain
\label{Madrid}}

\def\etal{et al.}
\newcommand{\Ha}{\hbox{{\rm H}$\alpha$}}
\newcommand{\Hb}{\hbox{{\rm H}$\beta$}}
\newcommand{\OII}{\hbox{[{\rm O}\kern 0.1em{\sc ii}]}}
\newcommand{\NeIII}{\hbox{[{\rm Ne}\kern 0.1em{\sc iii}]}}
\newcommand{\OIII}{\hbox{[{\rm O}\kern 0.1em{\sc iii}]}}
\newcommand{\NII}{\hbox{[{\rm N}\kern 0.1em{\sc ii}]}}
\newcommand{\SII}{\hbox{[{\rm S}\kern 0.1em{\sc ii}]}}
\newcommand{\HII}{\hbox{[{\rm H}\kern 0.1em{\sc ii}]}}

\begin{abstract}

  We present some of the first science data with the new Keck/MOSFIRE
  instrument to test the effectiveness of different AGN/SF diagnostics
  at $z \sim 1.5$.  MOSFIRE spectra were obtained in three $H$-band
  multi-slit masks in the GOODS-S field, resulting in two hour
  exposures of 36 emission-line galaxies.  We compare X-ray data with
  the traditional emission-line ratio diagnostics and the alternative
  mass-excitation and color-excitation diagrams, combining new MOSFIRE
  infrared data with previous HST/WFC3 infrared spectra (from the
  3D-HST survey) and multiwavelength photometry.  We demonstrate that
  a high $\OIII/\Hb$ ratio is insufficient as an AGN indicator at
  $z>1$.  For the four X-ray detected galaxies, the classic
  diagnostics ($\OIII/\Hb$ vs. $\NII/\Ha$ and $\SII/\Ha$) remain
  consistent with X-ray AGN/SF classification.  The X-ray data also
  suggest that ``composite'' galaxies (with intermediate AGN/SF
  classification) host bona-fide AGNs.  Nearly $\sim$2/3 of the $z
  \sim 1.5$ emission-line galaxies have nuclear activity detected by
  either X-rays or the classic diagnostics.  Compared to the X-ray and
  line ratio classifications, the mass-excitation method remains
  effective at $z>1$, but we show that the color-excitation method
  requires a new calibration to successfully identify AGNs at these
  redshifts.

\end{abstract}


\section{Introduction}

Every massive nearby galaxy hosts a supermassive black hole (SMBH),
and the mass of the SMBH correlates with the mass of the host galaxy
bulge \citep{mag98}.  Theoretical simulations suggest this connection
exists because past active galactic nucleus (AGN) phases of rapid SMBH
growth were associated with periods of massive star formation (SF) in
the host galaxy \citep[e.g.,][]{dim05,hop06}.  Observations of AGN
frequency, including both weak Seyferts and powerful quasars, in
different host galaxy types over the cosmic time can be used to
directly test models of coupled SMBH-galaxy growth.

Selection by blue optical color, X-ray emission, or infrared (IR)
color can be used to select powerful quasars to very high redshifts,
but these methods are less effective for finding obscured or
moderately-accreting AGNs.  Instead the most efficient way to find
moderate-luminosity AGNs is by their unique emission line signature.
Compared to typical star formation processes, the higher-ionization
radiation of an AGN tends to increase the ratios between rest-frame
optical collisionally excited ``forbidden'' lines and hydrogen
recombination lines \citep{bpt81,kew06}.  In particular the line
ratios $f(\OIII\lambda5007)/f(\Hb)$, $f(\NII\lambda6584)/f(\Ha)$, and
$f(\SII\lambda6718+6731)/f(\Ha)$ are typically used in the classic
``BPT'' and ``VO87'' diagnostics \citep{bpt81,vo87}: the wavelength
proximity of each line pair means the ratios are nearly insensitive to
reddening.  In the standard AGN unified model \citep{ant93} these
narrow emission lines can be detected even if the X-ray and
ultraviolet (UV) ionizing radiation source is absorbed by anisotropic
obscuration.  The AGN line ratio signature also remains visible for
SMBHs of moderately low accretion rates ($L/L_{Edd} \sim 10^{-3}$)
which are otherwise dominated by their host galaxy starlight
\citep[e.g.,][]{kau09}.

\begin{figure*}[ht]  
\begin{center}
  \epsscale{0.9}
  {\plotone{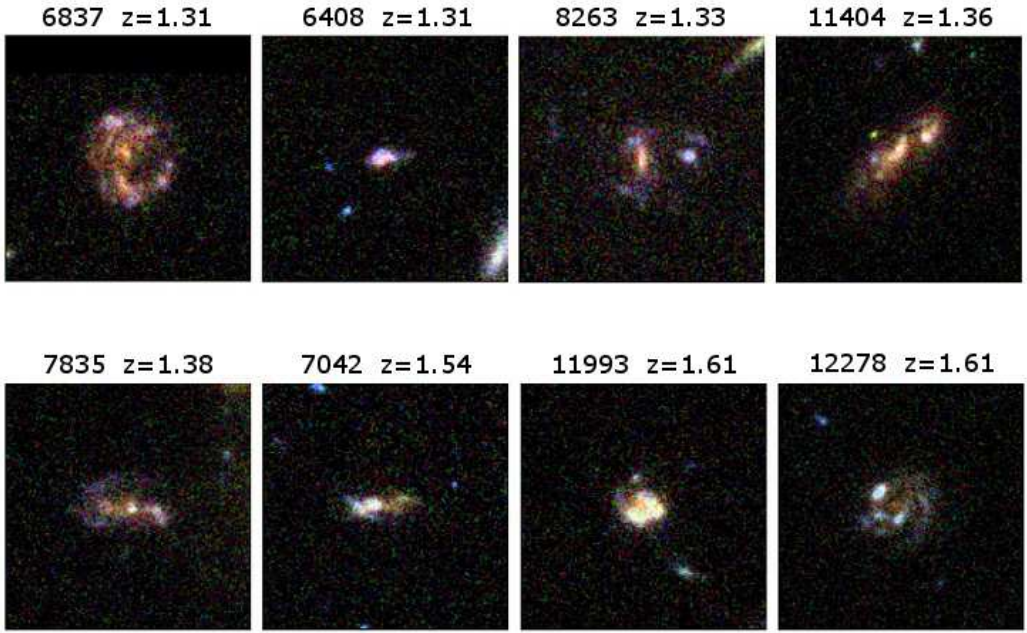}}
\end{center}
\figcaption{ACS $viz$ color-composite 5'' $\times$ 5'' images for 8 of
  the 36 emission-line galaxies, selected to be representative of the
  full sample's range in redshift, stellar mass, and AGN/SF
  classification.  At $z \sim 1.5$ these thumbnails span 42 $\times$
  42~kpc.  Galaxy \#11404 (upper right panel) is an X-ray AGN.  The
  rest-frame UV images of the full set of 36 galaxies (including those
  shown here) exhibit a range of disk-dominated and clumpy
  morphologies without strong point sources.
  \label{fig:images}}
\end{figure*}

There has been great success in using line ratio diagnostics to select
AGNs and characterize their hosts at $z \sim 0$
\citep[e.g.,][]{kau03b,hec04,yan06,schaw07}.  But extending a similar
AGN census to $z>0.4$ is difficult because the $\NII/\Ha$ and
$\SII/\Ha$ ratios redshift to the IR, where ground-based spectroscopy
has historically been expensive.  A high $\OIII/\Hb$ ratio alone is
degenerate between nuclear activity and metal-poor $\HII$ regions, and
the $\NII/\Ha$ or $\SII/\Ha$ line ratio is generally necessary to
distinguish AGNs from inactive galaxies with low metallicity.  Some
authors have suggested combining $\OIII/\Hb$ with bluer emission line
ratios, like $\OII\lambda3726+3729/\Hb$ \citep{lam04} or
$\NeIII\lambda3869/\OII$ \citep{tro11}.  However these bluer lines are
typically weaker, making them difficult to apply to distant galaxies.
Instead it is possible to exploit the correlation of metallicity with
color and stellar mass \citep{tre04} to altogether eliminate a second
line ratio.  These modified AGN/SF diagnostics use $\OIII/\Hb$ to
measure excitation, but replace the $\NII/\Ha$ ratio with rest-frame
color \citep{yan11} or stellar mass \citep{jun11}.  AGN selection
using the``color-excitation (CEx)'' and ``mass-excitation (MEx)''
methods agree well with the classic line ratio diagnostics at $z<0.4$
and X-ray selection at $z<1$.

It is less clear if the BPT, VO87, CEx, and MEx AGN/SF diagnostics are
directly applicable to higher redshift.  Star formation processes at
$z>1$ are generally different than in nearby galaxies, with higher gas
and dust fractions, younger stellar populations, lower metallicities,
and higher star formation rates
\citep[e.g.,][]{pap05,red06,tac09,shim11}.  Some authors argue that
the classical line ratio diagrams are unreliable at high redshift
because local starburst galaxies with SF rates typical of $z>1$
galaxies tend to have similarly high $\NII/\Ha$ and $\OIII/\Hb$ ratios
\citep{liu08,bri08}.

In this work we directly test the effectiveness of the classical line
ratio, color-excitation, and mass-excitation diagnostics for
identifying AGN at $z \sim 1.5$.  We use observations of 36 galaxies
at $1.30<z<1.62$ (mean $\bar{z}=1.52$) with the new MOSFIRE
multi-object spectrograph \citep{mclean10,mclean12} on the Keck
telescope.  The new MOSFIRE data are coupled to X-ray observations,
previous spectroscopy, and multiwavelength photometry for stellar
masses and rest-frame colors.

\section{Observational Data}

We study the emission line ratios, rest-frame colors, stellar masses,
and X-ray properties of 36 galaxies in the Great Observatories Origins
Deep Survey-South field \citep[GOODS-S,][]{goods}.  The targets were
selected from Hubble Space Telescope ({\it HST}) Wide Field Camera 3
(WFC3) G141 grism observations as part of the 3D-HST survey
\citep{3dhst}.  The initial selection required $F140W<24$ and a
detected $\OIII$ emission line in the redshift range $1.3<z<1.7$.  The
flux-limited sample includes a mix of star-forming, AGN, and composite
galaxies.  New Keck/MOSFIRE $H$-band spectra were obtained for the
$\NII/\Ha$ and $\SII/\Ha$ line ratios.  The GOODS-S field also
includes tremendously deep optical and IR photometry \citep{dah10} and
4~Ms of X-ray coverage \citep{xue11}.  Details on the WFC3 grism,
MOSFIRE, and photometric data are provided below.  Table
\ref{tbl:targets} includes the full suite of line ratios, colors, and
masses for the 36 galaxies.  ACS $viz$ color-composite images of eight
representative galaxies are shown in Figure \ref{fig:images}.  In
general the sample includes disk-dominated and clumpy morphologies
without strong point sources, similar to the clumpy galaxies and AGN
hosts of \citet{bou12}.

\subsection{Keck/MOSFIRE}

\begin{figure*}[ht] 
\epsscale{1.1}
{\plotone{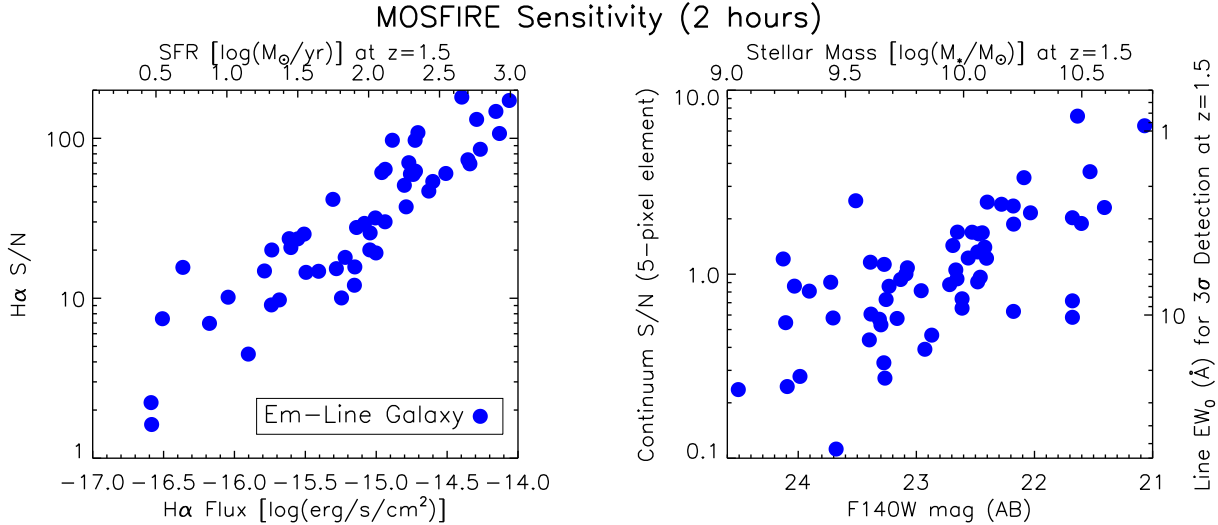}}
\figcaption{The sensitivity of MOSFIRE to emission lines and continuum
  measurements for two-hour exposures.  At left is shown the measured
  S/N and flux for $\Ha$ in our targets, with the upper axis
  translating these line fluxes to SFR (using the \citealt{ken98}
  relation at $z=1.5$ with a standard $\Lambda$CDM cosmology and
  $h_0=0.7$).  Note that the SFR axis is inaccurate for galaxies with
  significant AGN contribution to their $\Ha$ flux.  The right panel
  shows the continuum signal-to-noise (per 5-pixel resolution element)
  and corresponding rest-frame EW limit for 3$\sigma$ line detection
  versus $F140W$ magnitude and corresponding stellar mass, using the
  empirical relation $\log(M_*)=21.3-0.5m_{\rm F140W}$.
\label{fig:mosfiresn}}
\end{figure*}

MOSFIRE observations were performed Sept 14-15 and Oct 10 2012, with
three $H$-band masks in the GOODS-S field.  All targets were observed
in two dither positions within $0\farcs7$ slits, with total on-target
exposure times of two hours each.  The resulting wavelength coverage
is $1.46<\lambda<1.81$~$\mu$m with a spectral resolution of $R \sim
3200$ ($\sim$5$\AA$ per resolution element).  Spectra were reduced,
sky-subtracted, wavelength-calibrated, and 1D-extracted using the
public MOSFIRE data reduction pipeline\footnote{\tt
  http://code.google.com/p/mosfire/}.  The wavelength calibration was
very accurate, with shifts of $\Delta{z} \lesssim 0.001$ from previous
spectroscopic redshifts.  The spectra were not flux-calibrated: flux
calibration is unnecessary for ratios of closely separated emission
lines.

Figure \ref{fig:mosfiresn} shows the sensitivity of MOSFIRE, for both
emission lines and continuum measurements, in our two-hour exposures.
The left panel shows the emission-line signal-to-noise (S/N) with line
flux and star formation rate (SFR), using the \citet{ken98} relation
to convert line flux at $z=1.5$ to SFR.  The right panel shows the
continuum S/N (measured over a resolution element of 5 pixels) with
F140W magnitude from 3D-HST observations.  Continuum S/N is translated
to the minimum rest-frame equivalent width (EW$_0$) for a 3$\sigma$
line detection at $z=1.5$, and F140W magnitude is converted to stellar
mass using the relation $\log(M_*)=21.3-0.5m_{\rm F140W}$ (derived
using the stellar masses from SED fitting in Section 2.3).  In
general, our sensitivity measurements agree well with the estimates
provided on the MOSFIRE website\footnote{\tt
  http://irlab.astro.ucla.edu/mosfire/}.

\subsection{HST/WFC3 G141 Slitless Grism}

The GOODS-S field contains publicly available HST/WFC3 G141 grism
spectra to 2-orbit depth as part of the 3D-HST survey \citep{3dhst}.
We reduced the data using the aXe software \citep[][available at {\tt
    http://axe.stsci.edu/axe/}]{kum09}, producing 2D and 1D
wavelength- and flux-calibrated spectra with a wavelength range of
$1.1<\lambda<1.7\mu$m.  The grism resolution is $R \simeq 130$ for a
point source (46.5\AA/pixel) and somewhat worse for our extended
galaxies.  Spectra were inspected to identify contamination by
neighboring objects: all of the objects studied here have $\Hb$ and
$\OIII$ emission lines unaffected by contamination.

\subsection{Ancillary Photometry}

The GOODS-S field has well-sampled spectral energy distributions
(SEDs) from UV to IR wavelengths \citep{dah10}.  The deep photometry,
with high-confidence spectroscopic redshifts from MOSFIRE, allows for
robust estimates of rest-frame colors and stellar masses for $z=1.5$
galaxies.  These quantities are computed following the methods of the
Rainbow database\footnote{\tt
  https://rainbowx.fis.ucm.es/Rainbow\_Database/} \citep{barro11}.
First, the observed photometry is transformed to the rest-frame using
the spectroscopic redshift.  The rest-frame SED is then fitted to a
grid of \citet{bru03} models characterized by exponentially declining
star-formation histories, a \citet{cha03} initial mass function, and a
\citet{cal01} extinction law.  Stellar mass and rest-frame photometry
are measured from the best-fit template.  The well-sampled SED
guarantees that the rest-frame $U-B$ color is interpolated rather than
extrapolated.

\begin{figure*}[ht] 
\epsscale{1.15}
{\plotone{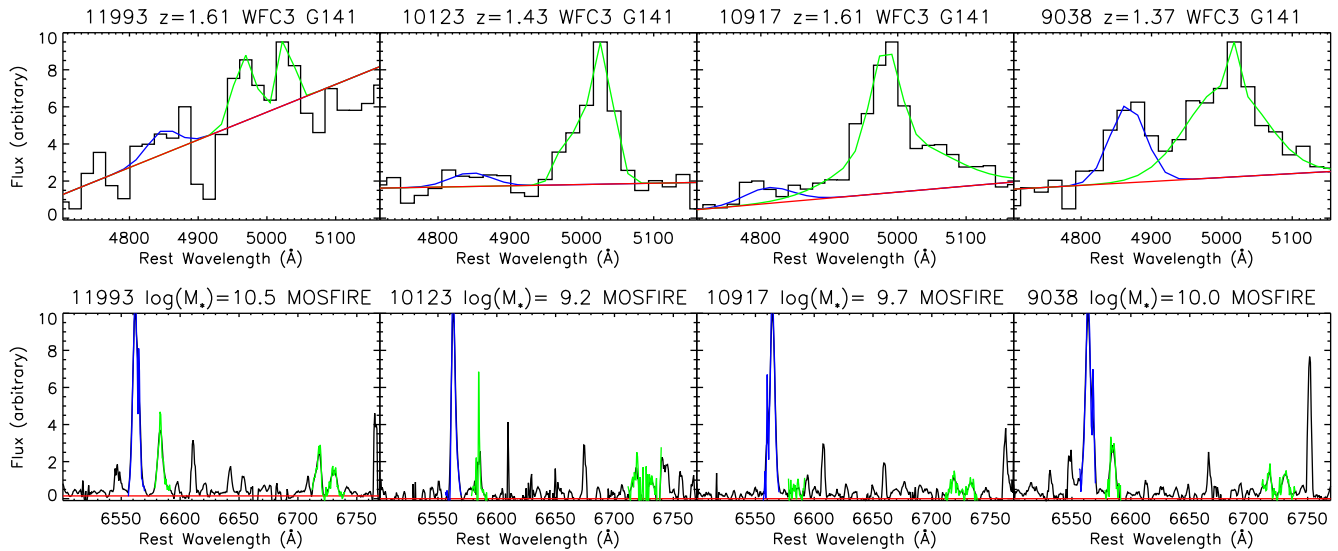}}
\figcaption{Examples of emission-line fits to the WFC3 grism and
  MOSFIRE spectra.  The black histogram shows the spectrum, with the
  continuum fit in red and the emission line regions in blue ($\Ha$
  and $\Hb$) and green ($\OIII$, $\NII$, and $\SII$).  In the
  high-resolution MOSFIRE data, the (uncalibrated) line intensity is
  computed by simply summing the continuum-subtracted spectrum in the
  line region.  However the $\Hb$ and $\OIII$ emission lines overlap
  in the low-resolution WFC3 grism data, and the line flux is instead
  calculated from Gaussians fitted to the continuum-subtracted
  spectrum.  The object in the right-most panels (ID \#9038) is X-ray
  detected, but has both emission line and X-ray properties of a
  star-forming galaxy.
\label{fig:linefits}}
\end{figure*}

The GOODS-S field also contains the deepest X-ray data in the sky,
with 4~Ms of {\it Chandra} observations \citep{xue11}.  Four of the 36
galaxies in this study are X-ray detected: three of these are hard
sources with $L_{\rm 2-8 keV} > 10^{42}$~erg/s and are classified as
AGNs by \citet{xue11}, while one has soft X-ray emission consistent
with star formation.  While the $L_x/SFR$ relation indicates that
$SFR>1000 M_\odot/yr$ can also produce $L_{\rm 2-8 keV} >
10^{42}$~erg/s \citep{leh10,min12}, such SFRs represent the extreme
upper end of our sample (Figure \ref{fig:mosfiresn}).  Indeed, all
three of the sources classified as AGNs have X-ray luminosities at
least 1.5~dex higher than expected by their $\Ha$-derived SFRs.  The
X-ray sources are marked in Table \ref{tbl:targets}, and we use the
three X-ray AGNs as ``truth'' for testing the line ratio AGN/SF
diagnostics in Section 4.

\section{AGN/SF Diagnostics}

We compare the classical line ratio with the color-excitation and
mass-excitation AGN/SF diagnostics, using emission line ratios
measured from the spectra and photometry-derived rest-frame colors and
stellar masses.  The MOSFIRE $H$-band spectra provide $\NII$, $\SII$,
and $\Ha$ fluxes, while $\Hb$ and $\OIII$ are measured from the WFC3
grism spectra.  Examples of the emission-line fitting are shown in
Figure \ref{fig:linefits}.

For the high-resolution MOSFIRE spectra, a continuum is fit across the
emission line regions by splining the 50-pixel smoothed continuum.
The emission line intensities are then computed simply as the sum of
the continuum-subtracted flux in each of the wavelength regions
$6556<\lambda<6570$ ($\Ha$), $6578<\lambda<6592$ ($\NII$), and
$6711<\lambda<6740$ (both $\SII$ lines).

Line flux measurements in the low-resolution WFC3 grism spectra are
more difficult because the $\Hb$ and $\OIII$ lines are blended.  For
the grism spectra we subtract a linear continuum over the wavelength
region $4750<\lambda<5120$ and then fit three Gaussians, each of which
are restricted to be within $20\AA$ ($\sim$1200 km~s$^{-1}$) of the
line centers ($4861\AA$, $4959\AA$, and $5007\AA$).  The
\OIII$\lambda$5007 flux is computed as $3/4$ of the sum of the blended
$\OIII$ Gaussians \citep{sto00}.

Errors in line fluxes and ratios are calculated by bootstrapping
10~000 realizations of the resampled data.  These errors accurately
quantify the difficulty in line fitting when there is significant
contamination by sky lines in MOSFIRE or nearby objects in the WFC3
grism.  We use $1\sigma$ upper limits for undetected emission lines.

\begin{figure*}[ht] 
\epsscale{1.15}
{\plotone{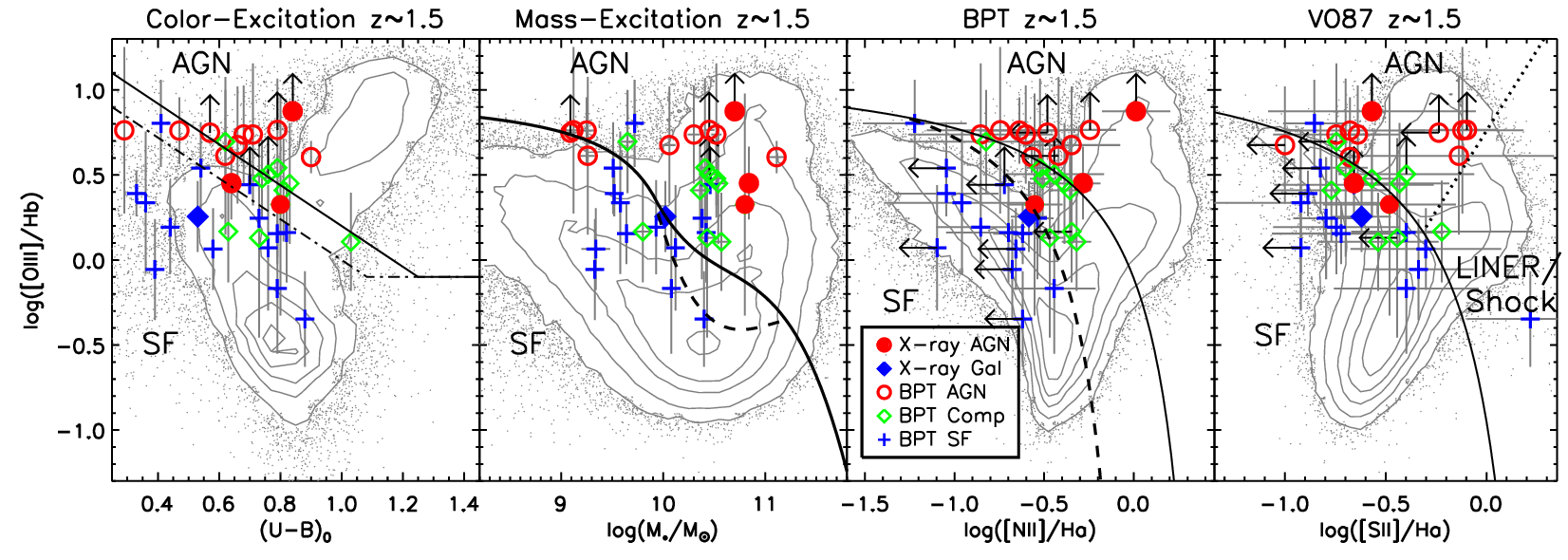}}
\figcaption{AGN/SF diagnostics: from left to right, color-excitation
  with $\OIII/\Hb$ flux ratio vs. rest-frame $U-B$ color
  \citep{yan11}, mass-excitation with $\OIII/\Hb$ vs. stellar mass
  \citep{jun11}, the $\OIII/\Hb$ vs. $\NII/\Ha$ BPT diagram, and the
  $\OIII/\Hb$ vs. $\SII/\Ha$ VO87 diagram.  The $\Hb$ and $\OIII$
  emission lines are measured from the WFC3 grism, while the $\Ha$,
  $\NII$, and $\SII$ lines are measured from the new MOSFIRE data.
  AGNs sit in the upper right of each panel, while star-forming
  galaxies (dominated by $\HII$ regions) sit below and to the left of
  the solid lines.  The dashed lines of the MEx and BPT diagrams
  display the ``composite'' regions of both nuclear activity and star
  formation, and the dotted line in the $\SII/\Ha$ panel separates
  Seyfert AGNs from LINERs.  Galaxies are color-coded based on their
  X-ray and \citet{kew06} classifications.  In contrast to $z<0.3$
  galaxies (gray contours), most of the $z \sim 1.5$ galaxies have
  properties near the AGN/SF dividing line of each diagnostic method.
  At $z \sim 1.5$, the MEx method remains consistent with the BPT
  diagnostics, but the CEx method misses several AGNs identified by
  the other methods.  The dot-dashed line in the CEx diagram shows our
  revised AGN/SF demarcation (see Section 4).
\label{fig:bpt}}
\end{figure*}

Figure \ref{fig:bpt} presents the color-excitation and mass-excitation
diagrams along with the traditional BPT ($\OIII/\Hb$ vs. $\NII/\Ha$
and $\OIII/\Hb$ vs. $\SII/\Ha$) AGN/SF diagnostics for the $z \sim
1.5$ galaxies.  The solid lines give the AGN/SF separation line
defined for each diagram (CEx: \citealt{yan11}; MEx: \citealt{jun11},
BPT and VO87: \citealt{kew06}), with AGNs having $\log(\OIII/\Hb)$
above this line and star-forming galaxies lying below.  Each panel
also includes a comparison sample of $z<0.3$ galaxies from the Sloan
Digital Sky Survey \citep[SDSS,][]{york00} MPA-JHU value-added DR7
catalog\footnote{{\tt http://www.mpa-garching.mpg.de/SDSS/}}.  These
$z<0.3$ galaxies have emission line and stellar mass measurements
described by \citet{tre04} and \citet{kau03a}, with rest-frame
magnitudes calculated using the {\tt kcorrect} IDL software
\citep{kcorrect}.  The $z \sim 1.5$ galaxies are color-coded according
to their \citet{kew06} AGN/SF classification as follows:

\begin{itemize}
  \item X-ray sources: filled red circles for AGNs, filled blue
    diamonds for galaxies \citep[as classified by][]{xue11}.
  \item SF galaxy, blue crosses:
    $\log(\OIII/\Hb)<1.3+0.61/[\log(\NII/\Ha)-0.05]$.
  \item Composite SF+AGN, open green diamonds:
    $\log(\OIII/\Hb)>1.3+0.61/[\log(\NII/\Ha)-0.05]$ and
    $\log(\OIII/\Hb)<1.19+0.61/[\log(\NII/\Ha)-0.47]$, from
    \citet{kau03b}.
  \item AGN, open red circles:
    $\log(\OIII/\Hb)>1.19+0.61/[\log(\NII/\Ha)-0.47]$ and
    $\log(\OIII/\Hb) > 0.76+1.89\log(\SII/\Ha)$.
  \item LINER/Shock: $\log(\OIII/\Hb)>1.19+0.61/[\log(\NII/\Ha)-0.47]$
    and $\log(\OIII/\Hb)<0.76+1.89\log(\SII/\Ha)$.
\end{itemize}

None of the $z \sim 1.5$ galaxies meet the LINER/Shock criteria,
although the elevated $\SII/\Ha$ ratios of several $z \sim 1.5$
galaxies classified as star-forming (by their $\NII/\Ha$ and
$\OIII/\Hb$ ratios) support the conclusion of \citet{yan12} that such
systems are excited by shocks rather than AGN ionization.

Galaxies at $z \sim 1.5$ have typically higher $\OIII/\Hb$ ratios than
$z<0.3$ galaxies, in agreement with previous $z>1$ studies
\citep[e.g.,][]{liu08,wri10,tru11}.  However this is not necessarily
because $z \sim 1.5$ galaxies are more likely to be AGNs:
low-metallicity star-forming galaxies (in the upper left of each panel
in Figure \ref{fig:bpt}) are also more likely at $z>1$.  An elevated
$\OIII/\Hb$ ratio is insufficient for AGN identification at $z>1$.

\section{AGN Identification at $z>1$}

Determining the effectiveness of each AGN/SF diagnostic requires
knowledge of the ``true'' AGN population.  The X-ray data provide an
independent AGN/SF classification method.  For the four X-ray detected
galaxies, the X-ray classifications of \citet{xue11} generally agree
with the \citet{kew06} classifications: the three X-ray AGNs have
emission line ratios indicative of either AGNs or composite AGN+SF
systems, while the X-ray galaxy has star-forming emission-line ratios.
The classic AGN/SF line ratio classifications remain accurate for the
four X-ray detected galaxies at $z \sim 1.5$.

The X-ray detection of galaxies in or near the ``composite'' region,
with line ratios intermediate between AGN and SF, implies that these
objects do in fact host accreting AGNs.  These data are in contrast
with \citet{liu08} and \citet{bri08}, who argued that AGN/SF composite
galaxies at $z>1$ are starbursts with unusual gas properties and no
significant AGN.  Galaxies in the composite region may host
starbursts, but the X-ray data imply that they also host genuine AGNs.
Indeed, many studies suggest a connection between star formation
activity and AGN accretion \citep[e.g.,][]{kau03b,tru12}.  The
analyses of \citet{wri10} and \citet{tru11}, with spatially-resolved
emission line ratio gradients, provide additional evidence that $z>1$
AGN/SF composite galaxies host nuclear activity.

The lack of an X-ray detection in the other 32 galaxies does not mean
that they lack AGNs.  X-rays surveys are less sensitive to
moderate-luminosity AGNs in galaxies of lower stellar masses
\citep{aird12}, and all three X-ray detected AGNs have the highest
stellar masses in our sample ($\log(M_*/M_\odot)>10.7$).  X-ray
surveys are also insensitive to heavily obscured AGNs.  Because the
\citet{kew06} classifications match the X-ray data for the high-mass
X-ray detected galaxies, we conclude that these classifications are
robust for $z>1$ galaxies.  Of the 32 X-ray undetected galaxies, those
identified as AGNs by their line ratios are presumably undetected
because they have low stellar masses or are X-ray obscured.

The total number of AGNs from the combined X-ray and line ratio
classifications is 11/36, with an additional 10/36 AGN/SF composite
galaxies.  Thus about 1/3 of our $z>1$ emission-line galaxies are
AGN-dominated, and nearly $\sim$2/3 have some AGN component.  This is
significantly higher than the $\sim$15\% AGN and composite fraction at
$z \sim 0$ \citep{kau03b}, although this may be partly driven by our
emission-line selection.

The mass-excitation method remains generally consistent with the
BPT/VO87 and X-ray classifications at $z \sim 1.5$.  \citet{jun11}
emphasize that a probabilistic approach is more meaningful with the
MEx method than simply classifying individual galaxies by the AGN/SF
dividing line.  Following \citet{jun11}, the MEx probability analysis
indicates that $\sim$18/36 galaxies are AGNs, matching the high AGN
fraction of the X-ray and BPT/VO87 classifications.

Meanwhile the color-excitation method, as calibrated at $z \sim 0$,
identifies only 7 of the 12 AGNs classified by X-rays or line ratios.
This is because a $z \sim 1.5$ galaxy has significantly higher SFR and
lower metallicity, and is consequently bluer, than a $z \sim 0$ galaxy
of the same mass \citep{erb06,man10}.  While the change in metallicity
is apparent in the increasing $\OIII/\Hb$ ratio, the change in star
formation rate is not.  Shifting the original AGN/SF division of
\citet{yan11} by 0.2 mags provides a much more effective separation
between AGNs from SF galaxies at $z \sim 1.5$.  Our data suggest the
following recalibrated color-excitation AGN/SF demarcation at $z>1$:

\begin{center}
  $\log(\OIII/\Hb) = {\rm max}[1.2-1.2(U-B)_0,-0.1]$
\end{center}


\section{Summary}

We use some of the first Keck/MOSFIRE data to test various diagnostics
of AGN/SF activity in 36 emission-line galaxies at $z \sim 1.5$.
Although only four sources are X-ray detected, their X-ray data
suggest that the classic emission-line ratio diagnostic remains
effective at $z>1$, and that ``composite'' galaxies (of intermediate
AGN/SF classification) do in fact host accreting AGNs.  We find that
nearly $\sim$2/3 of our $z \sim 1.5$ emission-line galaxies have some
AGN contribution detected by X-rays or line ratios.  Among alternative
AGN/SF diagnostics, the mass-excitation method remains consistent with
X-ray and line ratio classification at $z \sim 1.5$, but the
color-excitation method does not.  We suggest a recalibration for
AGN/SF classification with the color-excitation diagnostic at $z>1$.
A larger set of galaxies and emission line measurements will allow a
more detailed calibration and estimation of the AGN fraction at $z>1$.

\acknowledgements

JRT and the authors from UCSC acknowledge support from NASA HST grants
GO-12060.10-A and AR-12822.03, Chandra grant G08-9129A, and NSF grant
AST-0808133.  NPK acknowledges support from NSF grant AST-1106171.  We
owe tremendous gratitude to the MOSFIRE commissioning team for
development and support of a spectacular instrument.  We also thank
Greg Wirth, Marc Kassis, Jim Lyke, and the staff of Keck observatory
for excellent support while observing.  We wish to recognize and
acknowledge the very significant cultural role that the summit of
Mauna Kea has within the indigenous Hawaiian community: we are
fortunate to have the opportunity to conduct observations from this
mountain.

\begin{deluxetable*}{crrrrrrrr}
  \tablecolumns{9}
  \tablecaption{Galaxies Properties\label{tbl:targets}}
  \tablehead{
    \colhead{ID} & 
    \colhead{RA} & 
    \colhead{Dec} & 
    \colhead{$z$} & 
    \colhead{$f_{\OIII}/f_{\Hb}$\tablenotemark{a}} & 
    \colhead{$f_{\NII}/f_{\Ha}$\tablenotemark{b}} & 
    \colhead{$f_{\SII}/f_{\Ha}$\tablenotemark{b}} & 
    \colhead{$\log(M_*)$\tablenotemark{c}} &
    \colhead{$(U-B)_{\rm rest}$\tablenotemark{c}} \\
    \colhead{-} & 
    \colhead{(deg)} & 
    \colhead{(J2000)} & 
    \colhead{-} & 
    \colhead{-} & 
    \colhead{-} & 
    \colhead{-} & 
    \colhead{$(\log(M_{\odot}))$} &
    \colhead{(mag)} }
  \startdata
  8025 & 53.07059 & -27.75539 & 1.303 &  0.45 & $<$0.24 &  1.65 & 10.40 & 0.88 \\
  6837 & 53.12799 & -27.77140 & 1.306 & $>$5.82 &  0.57 &  0.80 & 10.45 & 0.79 \\
  9956 & 53.09090 & -27.73119 & 1.307 & $>$5.61 & $<$0.33 & $<$0.58 &  9.09 & 0.57 \\
  6408 & 53.13639 & -27.77499 & 1.308 &  1.16 & $<$0.22 &  0.50 &  9.34 & 0.58 \\
  11132 & 53.08729 & -27.71850 & 1.308 &  1.43 &  0.24 &  0.19 &  9.64 & 0.79 \\
  8263 & 53.02619 & -27.75230 & 1.327 &  1.47 & $<$0.45 &  0.60 &  9.80 & 0.63 \\
  11404\tablenotemark{d} & 53.12480 & -27.71710 & 1.356 &  2.12 &  0.28 &  0.33 & 10.80 & 0.80 \\
  11730 & 53.14540 & -27.71260 & 1.361 &  4.10 & $<$0.38 &  0.73 &  9.26 & 0.62 \\
  9038\tablenotemark{e} & 53.07529 & -27.74259 & 1.374 &  1.80 &  0.26 &  0.24 & 10.02 & 0.53 \\
  7835 & 53.05199 & -27.75839 & 1.376 & $>$2.77 & $<$0.19 &  0.22 & 10.46 & 0.70 \\
  10123 & 53.12810 & -27.72929 & 1.426 &  5.77 &  0.23 &  0.21 &  9.25 & 0.47 \\
  11010 & 53.05690 & -27.72030 & 1.473 &  1.56 &  0.14 &  0.18 &  9.93 & 0.44 \\
  7042 & 53.06560 & -27.76790 & 1.539 &  2.57 &  0.44 &  0.17 & 10.36 & 0.81 \\
  9915 & 53.13259 & -27.73229 & 1.549 &  1.28 &  0.48 &  0.29 & 10.57 & 1.03 \\
  11349 & 53.10960 & -27.71719 & 1.552 &  5.79 &  0.18 &  0.76 &  9.12 & 0.29 \\
  12961 & 53.09719 & -27.69860 & 1.576 &  1.18 & $<$0.08 & $<$0.12 & 10.12 & 0.76 \\
  5745 & 53.07379 & -27.78420 & 1.607 &  2.46 &  0.09 & $<$0.13 &  9.52 & 0.33 \\
  11993 & 53.11220 & -27.71100 & 1.608 &  3.01 &  0.31 &  0.27 & 10.52 & 0.74 \\
  13252 & 53.09469 & -27.69459 & 1.608 &  5.47 &  0.25 &  0.23 & 10.30 & 0.68 \\
  10366 & 53.07300 & -27.72680 & 1.608 &  4.97 &  0.15 & $<$0.18 &  9.65 & 0.62 \\
  8753 & 53.10430 & -27.74650 & 1.609 &  2.82 &  0.41 &  0.37 & 10.54 & 0.83 \\
  10917 & 53.10910 & -27.72150 & 1.610 &  6.36 &  0.06 &  0.14 &  9.72 & 0.41 \\
  6842 & 53.07300 & -27.77050 & 1.610 &  1.76 &  0.29 &  0.16 & 10.38 & 0.73 \\
  12278 & 53.16040 & -27.70770 & 1.610 &  1.35 &  0.34 & $<$0.36 & 10.43 & 0.73 \\
  11534 & 53.12319 & -27.71559 & 1.610 &  3.49 &  0.30 &  0.20 & 10.41 & 0.79 \\
  9792 & 53.03450 & -27.73340 & 1.611 &  3.47 & $<$0.09 & $<$0.15 &  9.51 & 0.54 \\
  10817\tablenotemark{e} & 53.12279 & -27.72279 & 1.612 & $>$7.49 &  1.03 &  0.27 & 10.70 & 0.84 \\
  8803 & 53.13290 & -27.74580 & 1.612 &  1.45 &  0.20 &  0.40 & 10.42 & 0.82 \\
  8154 & 53.14730 & -27.75349 & 1.612 & $>$3.20 &  0.36 &  0.40 & 10.45 & 0.76 \\
  7371 & 53.08409 & -27.76370 & 1.612 &  2.17 &  0.11 &  0.12 &  9.58 & 0.36 \\
  12285 & 53.11000 & -27.70789 & 1.613 &  5.47 &  0.14 &  0.18 & 10.52 & 0.71 \\
  7989 & 53.10279 & -27.75609 & 1.613 &  4.74 &  0.45 & $<$0.10 & 10.06 & 0.66 \\
  12703 & 53.11370 & -27.70149 & 1.613 &  4.03 &  0.27 &  0.21 & 11.11 & 0.90 \\
  12522\tablenotemark{e} & 53.10490 & -27.70520 & 1.613 &  2.83 &  0.52 &  0.22 & 10.84 & 0.64 \\
  8414 & 53.15650 & -27.75079 & 1.614 &  0.88 & $<$0.21 &  0.46 &  9.33 & 0.39 \\
  7499 & 53.14110 & -27.76189 & 1.622 &  0.68 &  0.36 &  0.40 & 10.08 & 0.79 \\
  \enddata

  \tablenotetext{a}{The $f_{\OIII}$ and $f_{\Hb}$ line flux measurements come from the WFC3 grism data.}
  \tablenotetext{b}{The $f_{\Ha}$, $f_{\NII}$, and $f_{\SII}$ emission line fluxes are measured from the MOSFIRE data.}
  \tablenotetext{b}{Stellar mass and rest-frame color are estimated from SED fitting, as described in Section 2.3.}
  \tablenotetext{d}{X-ray detected and classified as an AGN by \citet{xue11}.}
  \tablenotetext{e}{X-ray detected, but in the soft band only and
    consistent with emission from a star-forming galaxy \citep{xue11}.}
\end{deluxetable*}

\end{document}